\documentclass[twocolumn]{aastex62}

\usepackage{longtable}
\usepackage{graphicx}
\usepackage{amsmath,amssymb}
\usepackage{color}
\usepackage{units}
\usepackage{epstopdf}
\usepackage{hyperref}
\usepackage{multirow}
\usepackage{url}

\usepackage{subfigure}
\usepackage{rotating}

\newcommand{\beq}{\begin{equation}}
\newcommand{\eeq}{\end{equation}}
\newcommand{\bdm}{\begin{displaymath}}
\newcommand{\edm}{\end{displaymath}}

\definecolor{Gray}{gray}{0.9}
\definecolor{orange}{rgb}{0.9,0.5,0}

\graphicspath{{./plots/}}

\begin{document}

\title{2900 square degree search for the optical counterpart of short gamma-ray burst GRB 180523B with the Zwicky Transient Facility}

\author[0000-0002-8262-2924]{Michael W. Coughlin}
\affil{Division of Physics, Math, and Astronomy, California Institute of Technology, Pasadena, CA 91125, USA}

\author[0000-0002-2184-6430]{Tom{\'a}s Ahumada}
\affil{Department of Astronomy, University of Maryland, College Park, MD 20742, USA}

\author[0000-0003-1673-970X]{S. Bradley Cenko}
\affil{Astrophysics Science Division, NASA Goddard Space Flight Center, Mail Code 661, Greenbelt, MD 20771, USA}
\affil{Joint Space-Science Institute, University of Maryland, College Park, MD 20742, USA}

\author{Virginia Cunningham}
\affil{Department of Astronomy, University of Maryland, College Park, MD 20742, USA}

\author[0000-0001-9901-6253]{Shaon Ghosh}
\affil{Department of Physics, University of Wisconsin-Milwaukee, Milwaukee, WI 53211, USA}

\author[0000-0001-9898-5597]{Leo P. Singer}
\affil{Astrophysics Science Division, NASA Goddard Space Flight Center, Mail Code 661, Greenbelt, MD 20771, USA}

\author[0000-0001-8018-5348]{Eric C. Bellm}
\affil{Department of Astronomy, University of Washington, Seattle, WA 98195, USA}

\author{Eric Burns}
\affil{Astrophysics Science Division, NASA Goddard Space Flight Center, Mail Code 661, Greenbelt, MD 20771, USA}

\author{Kishalay De}
\affil{Division of Physics, Math, and Astronomy, California Institute of Technology, Pasadena, CA 91125, USA}

\author{Adam Goldstein}
\affil{Science and Technology Institute, Universities Space Research Association, Huntsville, AL 35805, USA}

\author[0000-0001-8205-2506]{V. Zach Golkhou} 
\affiliation{
DIRAC Institute, Department of Astronomy, University of Washington, 3910 15th Avenue NE, Seattle, WA 98195, USA}
\affiliation{The eScience Institute, University of Washington, Seattle, WA 98195, USA}
\altaffiliation{Moore-Sloan, WRF Innovation in Data Science, and DIRAC Fellow.}

\author[0000-0001-6295-2881]{David~L.\ Kaplan}
\affil{Department of Physics, University of Wisconsin-Milwaukee, Milwaukee, WI 53211, USA}

\author{Mansi M. Kasliwal}
\affil{Division of Physics, Math, and Astronomy, California Institute of Technology, Pasadena, CA 91125, USA}

\author{Daniel A. Perley}
\affil{Astrophysics Research Institute, Liverpool John Moores University, IC2, Liverpool Science Park, 146 Brownlow Hill, Liverpool L3 5RF, UK}

\author{Jesper Sollerman}
\affil{The Oskar Klein Centre, Department of Astronomy, 
AlbaNova, SE-106 91 Stockholm , Sweden}

\author{Ashot Bagdasaryan}
\affiliation{Caltech Optical Observatories, California Institute of Technology, Pasadena, CA 91125, USA}

\author{Richard G. Dekany}
\affiliation{Caltech Optical Observatories, California Institute of Technology, Pasadena, CA 91125, USA}

\author[0000-0001-5060-8733]{Dmitry A. Duev}
\affiliation{Division of Physics, Math, and Astronomy, California Institute of Technology, Pasadena, CA 91125, USA}

\author{Michael Feeney}
\affiliation{Caltech Optical Observatories, California Institute of Technology, Pasadena, CA 91125, USA}

\author{Matthew J. Graham}
\affiliation{Division of Physics, Mathematics and Astronomy, California Institute of Technology, Pasadena, CA 91125, USA}

\author{David Hale}
\affiliation{Caltech Optical Observatories, California Institute of Technology, Pasadena, CA 91125, USA}

\author[0000-0001-5390-8563]{Shri R. Kulkarni}
\affiliation{Division of Physics, Math, and Astronomy, California Institute of Technology, Pasadena, CA 91125, USA}
\affiliation{Caltech Optical Observatories, California Institute of Technology, Pasadena, CA 91125, USA}

\author{Thomas Kupfer}
\affiliation{Kavli Institute for Theoretical Physics, University of California, Santa Barbara, CA 93106, USA}
\affiliation{Department of Physics, University of California, Santa Barbara, CA 93106, USA}

\author{Russ R. Laher}
\affiliation{IPAC, California Institute of Technology, 1200 E. California Blvd, Pasadena, CA 91125, USA}

\author[0000-0003-2242-0244]{Ashish Mahabal}
\affiliation{Division of Physics, Mathematics, and Astronomy, California Institute of Technology, Pasadena, CA 91125, USA}
\affiliation{Center for Data Driven Discovery, California Institute of Technology, Pasadena, CA 91125, USA}

\author[0000-0002-8532-9395]{Frank J. Masci}
\affiliation{IPAC, California Institute of Technology, 1200 E. California Blvd, Pasadena, CA 91125, USA}

\author[0000-0001-9515-478X]{Adam A. Miller}
\affiliation{Center for Interdisciplinary Exploration and Research in Astrophysics and Department of Physics and Astronomy, Northwestern University, 2145 Sheridan Road, Evanston, IL 60208, USA}
\affiliation{The Adler Planetarium, Chicago, IL 60605, USA}

\author{James D. Neill}
\affiliation{Division of Physics, Math, and Astronomy, California Institute of Technology, Pasadena, CA 91125, USA}

\author[0000-0002-4753-3387]{Maria T Patterson}
\affiliation{DIRAC Institute, Department of Astronomy, University of Washington, 3910 15th Avenue NE, Seattle, WA 98195, USA}

\author{Reed Riddle}
\affiliation{Caltech Optical Observatories, California Institute of Technology, Pasadena, CA 91125, USA}

\author[0000-0001-7648-4142]{Ben Rusholme}
\affiliation{IPAC, California Institute of Technology, 1200 E. California Blvd, Pasadena, CA 91125, USA}

\author{Roger Smith}
\affiliation{Caltech Optical Observatories, California Institute of Technology, Pasadena, CA 91125, USA}

\author{Yutaro Tachibana}
\affiliation{Tokyo Institute of Technology, 2-12-1 Ookayama, Meguro, Tokyo 152-8551, Japan}

\author{Richard Walters}
\affiliation{Caltech Optical Observatories, California Institute of Technology, Pasadena, CA 91125, USA}

\begin{abstract}

There is significant interest in the models for production of short gamma-ray bursts.
Until now, the number of known short gamma-ray bursts with multi-wavelength afterglows has been small. While the {\it Fermi} Gamma-Ray Burst Monitor detects many gamma-ray bursts relative to the Neil Gehrels {\it Swift} Observatory, the large localization regions makes the search for counterparts difficult. 
With the Zwicky Transient Facility recently achieving first light, it is now fruitful to use its combination of depth ($m_\textrm{AB} \sim 20.6$), field of view ($\approx$ 47 square degrees), and survey cadence (every $\sim 3$ days) to perform Target of Opportunity observations.
We demonstrate this capability on GRB 180523B, which was recently announced by the {\it Fermi} Gamma-Ray Burst Monitor as a short gamma-ray burst.
ZTF imaged $\approx$ 2900\,square degrees of the localization region, resulting in the coverage of 61.6\,\% of the enclosed probability over 2 nights to a depth of $m_\textrm{AB} \sim 20.5$.
We characterized 14 previously unidentified transients, and none were found to be consistent with a short gamma-ray burst counterpart.
This search with the Zwicky Transient Facility shows it is an efficient camera for searching for coarsely-localized short gamma-ray burst and gravitational-wave counterparts, allowing for a sensitive search with minimal interruption to its nominal cadence. 
\end{abstract}

\section{Introduction}

Gamma-ray bursts (GRBs), 50 years after their discovery by the Vela Satellites \citep{KlSt1973}, are still among the brightest observable objects in the universe. With their intrinsic brightness and discovery at cosmological distances \citep{MeDj1997,GeMe2012}, they remain a mystery and an active area of research. 
Even their classification into the traditional ``short'' (SGRB) and ``long'' (LGRB) classes is subject to debate \citep{NoBo2006,ZhCh2008,BrNa2013}.
There are a variety of models for GRBs that have been developed since their discovery. For example, the ``fireball'' model \citep{WiRe1997,MeRe1998}, where a highly relativistic jet of electron/positron/baryon plasma is emitted by a compact central engine produced from the merger of two neutron stars (or one neutron star and one black hole), predicts production of gamma rays and hard X-rays within the jet. In addition, an ``afterglow'' is produced by interaction of the jet with nearby ambient material, resulting in emission in the X-ray, optical, and radio for several days following the GRB. Later models have been produced to understand observed deviations in lightcurves seen from this model \citep{WiOb2007,CaGe2009,MeGi2011,DuMa2015}, including slow-moving cocoons \citep{NaHo2014,LaLo2017,KaNa2017,MoNa2017} and Gaussian structured
jets \citep{KuGr2003,AbEA2017e,TrPi2017}.
In the case of a compact merger, there is an additional component of highly neutron rich, unbound matter which is driven by radioactive decay of r-process elements that can heat the ejecta and power a thermal ultraviolet/optical/near infrared transient known as a \emph{kilonova} (or \emph{macronova}) \citep{LaSc1974,LiPa1998,MeMa2010,Ro2015,KaMe2017}.
These sources are expected to be broadly isotropic and therefore it may be possible to observe a kilonova regardless of the orientation of the system \citep{RoKa2011}.

With both its depth ($m_\textrm{AB} \sim 20.6$ in 30\,s), wide field of view (FOV) ($\approx$ 47 square degrees), and rapid cadence (every $\sim 3$ days), the Zwicky Transient Facility (ZTF) \citep{Bellm2018,Graham2018,DeSm2018}, a camera and associated observing system on the Palomar 48 inch telescope, is important for coordinated follow-up of short GRB events.
The cadence in particular is important for follow-up, as the previous imaging provides significant constraints on the explosion time of any new transients.
It improves significantly in FOV over the Palomar Transient Factory \citep{LaKu2009,RaKu2009}, which had a 7.3 square degree imager capable of performing $\approx 100$ degrees squared searches \citep{SiCe2013}, and the regular cadence of observations places stringent limits on the explosion time of new transients.
It supplements existing systems like the Gravitational-wave Optical Transient Observer (GOTO) \citep{Ob2018} and Asteroid Terrestrial-impact Last Alert System (ATLAS) \citep{ToDe2018}.
It uses ``reference images,'' which are stacks of typically at least 15 images \citep{MaLa2018}.
These reference images are created for a fixed grid of tiles with minimal dithering, which is due to the desire to simplify the data processing for image subtraction \citep{Bellm2018}.
At the time of paper submission (November 2018), about 70\% of the sky visible from Palomar currently has reference image coverage available in g- and r-bands; the exact numbers differ due to the varying number of observations in each filter.
These references images are required for image differencing, which uses the ZOGY algorithm \citep{ZaOf2016} to identify moving and changing sources.

ZTF is part of the Global Relay of Observatories Watching Transients Happen (GROWTH) network of telescopes for rapid classification of identified candidates, whose follow-up is coordinated via a marshal developed for this purpose \citep{Kasliwal2018}. 
The idea of the marshal is that it allows time-domain astronomers to create filters to save sources from different discovery streams, such as ZTF, and coordinate follow-up with various queue scheduled or human operated telescopes.

GRB 180523B triggered the {\it Fermi} Gamma-ray Burst Monitor at 18:46:28.11 (UTC) on 2018 May 23. It was classified as a GRB by the flight software and localized by the automated system which performs the automated production and distribution of the localizations. Its localization region had a 1-sigma circular-equivalent statistical error of 13 degrees (including the systematic error results in a 90\% probability region $\approx 6600\,$square degrees). It was automatically flagged as a likely SGRB. The GRB was given a GBM trigger number, 180523782, and the standard analysis is reported in the Fermi GBM online burst catalog \citep{BhMe2016}. The duration, calculated as the usual $T_{90}$, is 1.984\,s.  While this is near the historical split time of 2\,s, modeling the GBM $T_{90}$ distribution as two log-normal functions yields a 90\% probability that this GRB belongs to the short duration class. The time-integrated best fit model is a comptonized model (a power law with an exponential cutoff), which is typical for SGRBs. The peak energy is $\approx$\,1\,MeV (1150 $\pm$ 711\,keV), which is in the top quartile for SGRBs, lending additional evidence to the classification of a short hard GRB. The burst has a 64\,ms peak flux at the 20th percentile for SGRBs and a fluence at the 40th percentile. The prompt brightness suggests the afterglow is likely to be of average luminosity \citep{NyFr2009}.

We are interested in systematically detecting the optical afterglow of SGRBs to characterize their emission mechanisms and their hosts. 
This is a part of a broader Target of Opportunity (ToO) program searching for optical counterparts to neutrinos and gravitational waves as well.
This program follows a Palomar Transient Factory program searching for LGRB afterglows from the {\it Fermi} Gamma-ray Burst Monitor; 8 afterglows were found \citep{SiCe2013,SiKa2015}.
Only a few other LGRB afterglows have been discovered serendipitously, e.g. \cite{StTo2017}.
These data will supplement the existing catalog of lightcurves from many other SGRBs \citep{TaLe2013,BeWo2013,SoBe2006,ZhZh2007,AnDa2009,ZhZh2015,FoFr2005,LeTa2006,McFo2008,AcAs2010,TrSa2016,BePr2005,RoBe2006,PeMe2009,FoMa2016,AbEA2017e}.
The multi-wavelength data is essential for constraining redshifts, energetics, emission geometry, and conditions in the circumburst medium, while their locations within galaxies provide clues as to their progenitors \citep{FoBe2013}.
In this paper, we describe the ToO search for the optical counterpart of GRB 180523B.
We describe our observing plan in Section~\ref{sec:observing}.
We summarize image quality information in Section~\ref{sec:images}.
The identified candidates, including their follow-up, are summarized in Section~\ref{sec:candidates}.
Section~\ref{sec:conclusions} summarizes our conclusions and future outlook.

\section{Observing Plan}
\label{sec:observing}

\begin{figure}[t]
 \includegraphics[width=3.5in]{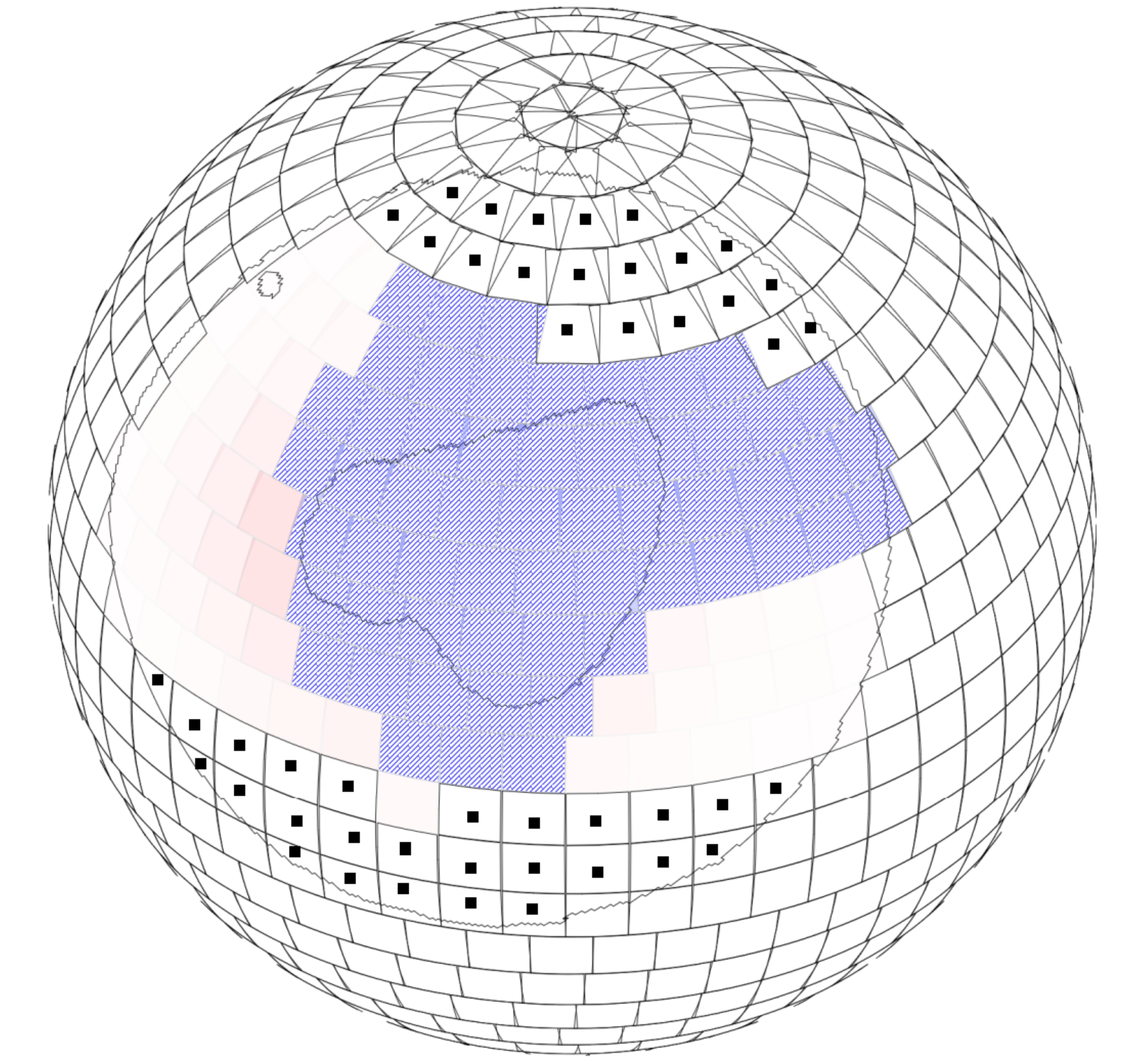}
  \caption{Target of Opportunity marshal representation of GRB 180523B scheduling. The black contours indicate the 50\% and 90\% credible levels. Tiles within the 90\% credible interval are considered for imaging. The unshaded tiles with a single dot within the 90\% credible interval do not contain reference images and therefore are not eligible to be analyzed for new transients with difference imaging. The red tiles shown contain reference images, but were not imaged due to time, airmass, or moon constraints. The checkered blue tiles shown are those scheduled to be imaged. Table \ref{table:images} shows the success rate of the requested observations.}
 \label{fig:ToO}
\end{figure} 

The observing strategy was as follows.  The fields we first observed corresponded to those enclosed in the 90\% localization region (see Figure~\ref{fig:ToO}). On the first night (2018 May 24), we planned a series of r-band, g-band, and then again r-band images.
Each set of observations began immediately after the first; this sequence allowed for measurement of both color and fade in any transients detected, useful for their eventual classification. 
We note that this left more than half of the night for the regular survey mode. 
To determine which fields would be imaged, we removed any that would not be available at an airmass less than 3 during the portion of the night when the localization region was visible from Palomar. 
The scheduling web interface includes airmass plots as well to help human observers determine when to schedule observations for.
In addition, we removed any fields without reference images in the bands of interest. From there, we scheduled 60\,s exposures in r-band, g-band, and then r-band, which are twice as long as the nominal 30\,s exposures performed in survey mode. We ordered the observations by right ascension. 
We note that while the tiles were selected using an algorithm, they were scheduled manually by calling the on-duty astronomer.
This schedule consists of an ordered list of sky locations and exposure times for the desired images.
We obtained images covering $\approx$ 2900 square degrees beginning at 3:51 UT on 2018 May 24, which is about 9.1\,hours after the burst trigger time. This delay was due to the fact that the SGRB alert came during the day at Palomar, and 3:51 UT was the start of nightly observations.

On the second night, we took a series of r-band and g-band exposures in the same fields as the night before. This time, 75\,s exposures were scheduled for the same time as the night before in order to be more sensitive and catch any potentially fading counterparts.
In total, we used 2:16 hours on the first night and 2:43 hours on the second night.
For comparison, ZTF observed for about 7:50 hours those nights, and so we used $\approx$\,25\% of the available observation time.

The ZTF FOV is comprised of 16 CCDs, each composed of 4 quadrants, placed in a $4 \times 4$ orthogonal grid, with about 12\% of the nominal tile area lost to gaps between the detectors (known as chip gaps). 
The generation of the observation schedule initially ignored the presence of the chip gaps. The total source-localization probability enclosed on 2018 May 24, accounting for the loss of probability due to the chip gaps, is 61.6\%. In the second night of observations (2018 May 25), where two observation windows in a single night was used, the total sky-localization probability enclosed in the observed fields was 68.4\%. The slightly greater containment of probability is due to the different airmass constraint requirements used in the first observation, where three observation windows were required. We also note that the order of the tiles by probability does not change with the inclusion of the chip gaps.

In addition to candidates that had the expected number of exposures over the two nights (5 if they were in a g and r field, 3 in just r, 2 in just g), some objects appeared in more than one field. In some of those cases, the objects were near the edges of the chips and were not found in those observations. There were also some objects for which there were not the expected number of exposures (only 1 in g, etc.). Once again, some of these candidates fall in chip gaps, which move from image to image.

\section{Image Quality}
\label{sec:images}

Successful analysis of a requested ZTF observation is a two-step process. Requested fields must successfully pass through both the ZTF robotic-observing program and also through the image-processing pipeline at the Infrared Processing and Analysis Center (IPAC). Therefore, the ``success rate'' is the percentage of quandrants with requested observations were processed.

We show the success rate in the entire image processing pipeline in Table \ref{table:images}. 108 fields (with 6,912 total quadrants), were requested for observation on the first night (2018 May 24). Of those 108 fields, 105 were successfully processed by IPAC. Some quadrants fail the processing due to bad seeing ($>$5''). $12\%$ of the quadrants failed processing for reasons of this type. 95 fields (6,080 total quadrants) were requested on the second night (2018 May 25). Of those 95 fields, 94 were successfully processed by IPAC, with only 2\% of quadrants failing processing. As weather conditions are the dominant contribution to missing quadrants, there is not much gain to be expected. The success rate over the course of both nights is 92.8\,\%.

We can estimate the loss of probability due to these quadrant failures.
Assuming that the scales in which the sky-localization changes is much larger than the quadrant size, we can estimate the total loss of probability enclosed as 7.4\% and 4.9\,\% on night 1 and 2 respectively.
The 5\,$\sigma$ median depth of the observations over the two nights is $m_\textrm{AB} > 20.3$ in r-band and $m_\textrm{AB} > 20.6$ in g-band. 

\begin{table}[ht]
\centering
\caption{Success rates for the requested ZTF observation fields over each night. The percentage of fields processed includes those with any quadrant in that field is processed.}
\label{table:images}
\begin{tabular}{|c|c|c|c|c|c|}
\hline
Date & Fields & Fields & Fields & Total Quadrant \\
 & Requested & Observed & processed & Success Rate \\ \hline
24-May & 108 & 106 & 105 & 88\% \\ \hline
25-May & 95 & 94 & 94 & 98\% \\ \hline
\end{tabular}
\end{table}

\section{Candidates}
\label{sec:candidates}

\subsection{Detection and filtering}
Once the observations are taken and processed by the IPAC pipeline, alerts known as AVRO packets are created for each object \citep{MaLa2018,Patterson2018}. These alerts contain information about the transient, including its magnitude, proximity to other sources and its previous history of detections among other metrics. 
The GROWTH collaboration has developed a framework that can handle the large number of alerts that ZTF produces in order to select the objects that are potentially interesting. Moreover, as each group within the GROWTH collaboration has a particular scientific focus, the language provides a tool to filter, discriminate and retrieve object of interest \citep{Kasliwal2018}. 

\begin{figure}[t]
 \begin{center}
\includegraphics[width=.5\textwidth]{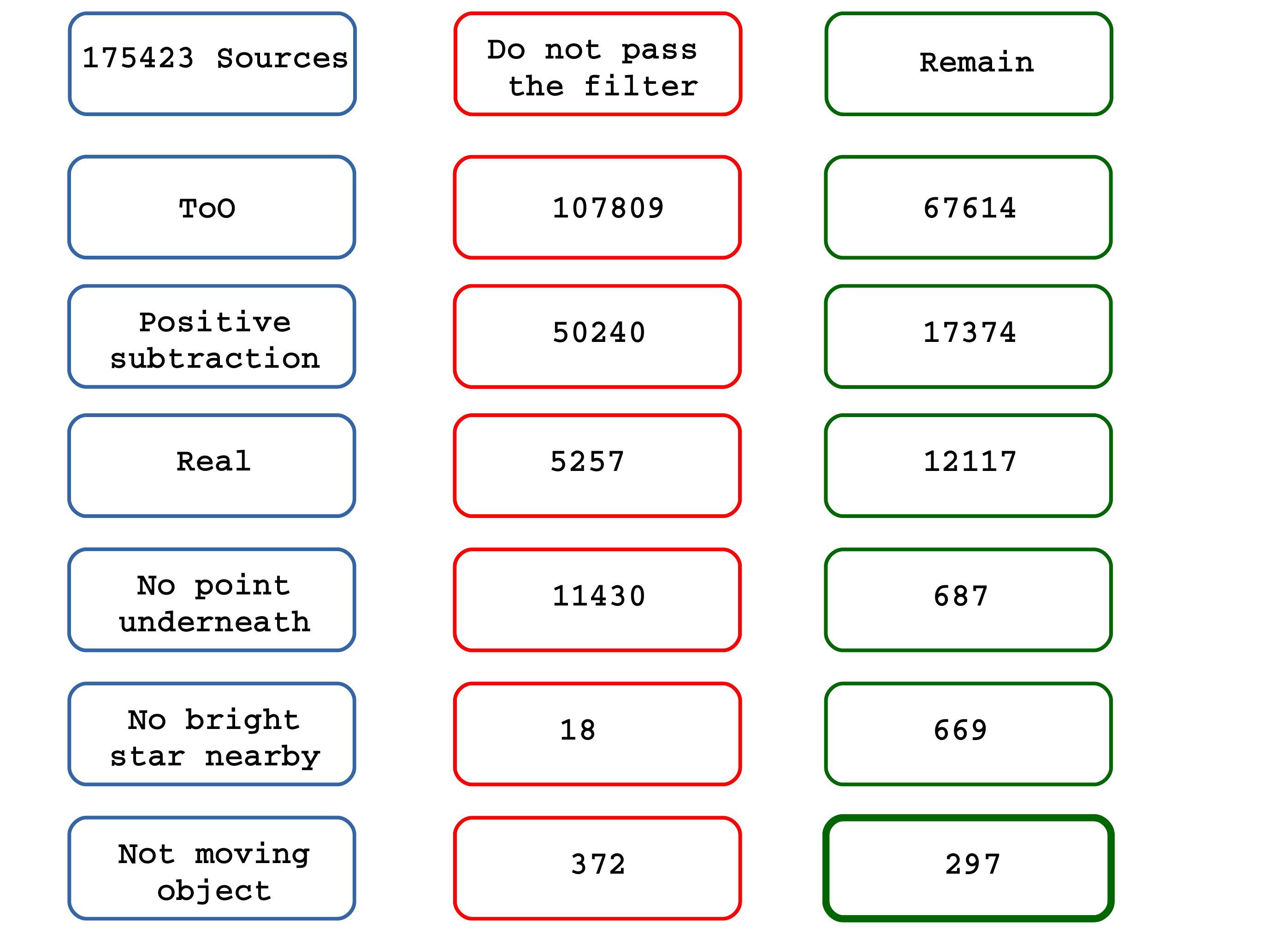}
\end{center}
\caption{Flow diagram of the filtering scheme for candidate transient sources detected after difference imaging. Each step is described in Section 4.2. 
The number of transients that survived after all the filter steps is 297 for the night of May 25th. Objects with previous detections are still included for scanning.}
 \label{fig:flowchart}
\end{figure}

In particular, for looking for optical counterparts of SGRBs, the filter restricted candidates to those with the following properties (see Figure~\ref{fig:flowchart}):
\begin{itemize}
\item  \textbf{Part of the ToO:} The candidates must belong to the fields related to the ToO observation of GRB 180523B.
\item  \textbf{Positive subtraction:} As the transients from ZTF images are detected by comparing reference images to the most recent image, it is required that the subtraction yields a positive difference in order to be considered a valid candidate. This is in contrast to negative subtractions, which are sources that decreased in brightness since the reference images were taken.
\item  \textbf{It is real:} The GROWTH collaboration has developed a Real-Bogus index \citep{Mahabal2018} trained on common image artifacts, including hot pixels and ghosts from bright stars. The index is the result of a random forest classifier and its value was restricted to $> 0.15$, the optimal to determine whether the source is a product of an artifact or if it is real.
\item \textbf{No point source underneath:} The coordinates of the source should not coincide with a point source \citep{TaMi2018}. This helps to rule out stellar variable sources or extragalactic transient/variable sources that are at small offsets from their hosts.
\item  \textbf{Two detections:} The object should have at least two detections during the night; asteroids, other solar system objects, and cosmic rays should not have a previous detection. 
\item  \textbf{Far from a bright star:} To avoid any kind of artifacts due to bright nearby source, including diffraction spikes or ghosts, the candidate is rejected if it is in the vicinity (20 arcsec) of a bright star ($m_\textrm{AB} < 15$).
\end{itemize}

\subsection{Scanning and selection}

During each night, the data are filtered, and the results are shown in the marshal for each scientific program. 
The interface shows the annotations retrieved from the filter, including color, cross-matches to external catalogs and positioning of the object on the chip and focal plane, as well as the historical light-curve of the object and subtracted images. This allows an on-duty astronomer to visually inspect the results of the marshal filter and proceed with a more refined candidate selection by saving the most interesting candidates into the project's report.
The interface allows individuals to cross-match the candidates with a number of databases in order to rule out candidates based on the available data. For example, it is possible to rule out a candidate based on WISE colors \citep{WrEi2010}, since AGNs and quasars lay in a particular locus in the WISE color-space \citep{StAs2012}. Likewise, a candidate can be ruled out if it has an ATLAS \citep{ToDe2018} detection registered before the SGRB alert. 
While we do not explicitly use the $g-$ and $r-$ band colors in the filtering, they are also potentially useful for classifying objects.

A number of candidates in the GROWTH report page presented a pattern of consecutive detection and non-detection in very short periods of time, thus requiring further investigation; these were determined to be artifacts from the ghosts of bright stars.
In addition, we do not consider sources that were first detected by ZTF prior to the time of the SGRB.
This resulted in 29 sources, 15 of which were artifacts; different scenarios can cause the presence of a bogus source, such as a nearby bright source or a bad pixel in the CCD.
Thus, for GRB 180523B the final candidate selection has 14 objects, whose properties are noted in Table~\ref{table:followup}; 11 of these have been followed-up, as described in the following subsection.

\subsection{Follow-up and results}

For the candidates that met the requirements, we requested additional photometry and spectroscopy. However, the observability of some of the transients was airmass constrained or required observations soon after sunset. This resulted in 3 candidates not being observed; these objects are noted at the bottom of Table~\ref{table:followup}.

The GROWTH collaboration has access to a number of facilities to follow-up the candidates. Particularly, photometric data were obtained using the Kitt Peak EMCCD Demonstrator (KPED) on the Kitt Peak 84 inch telescope \citep{Coughlin2018}. This instrument has the ability to take a set of short exposures and to modify its gain within a range, but the conventional long exposure-fixed gain mode was used for this follow-up. The photometric data taken with KPED was dark subtracted and flat field calibrated, and the magnitudes were determined using Source Extractor. The spectroscopic data were obtained using both the SED Machine (SEDM) \citep{nblago18} on the Palomar 60 inch (P60) telescope and the Double Spectrograph (DBSP) on the Palomar 200 inch (P200) telescope.

The classification for the candidates that had spectra was rapid and resulted in four new supernova Ia discovered.
In Table~\ref{table:followup}, there is a summary of the data of the candidates.
As previously mentioned, most of the candidates set very early in the night, reaching airmasses of 2 at the beginning of the observing run. 
3 of these candidates were not observable by follow-up facilities due to airmass constraints.
However, 7 of these objects were imaged with KPED a few weeks after their discovery. None of the sources had faded completely, indicating that they are not related to a SGRB. 

\begin{figure*}[t]
 \begin{center}
\includegraphics[width=.49\textwidth]{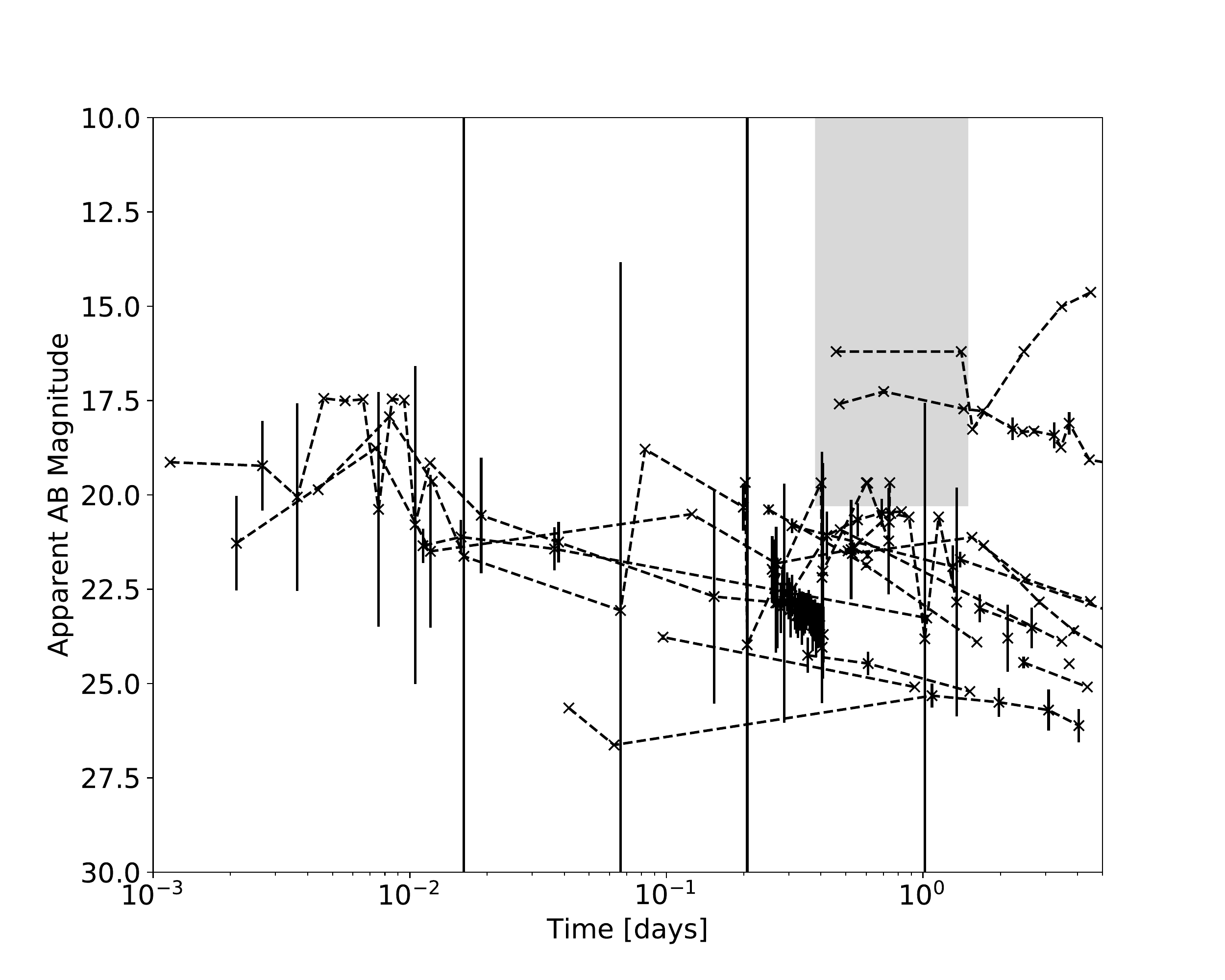}
\includegraphics[width=.49\textwidth]{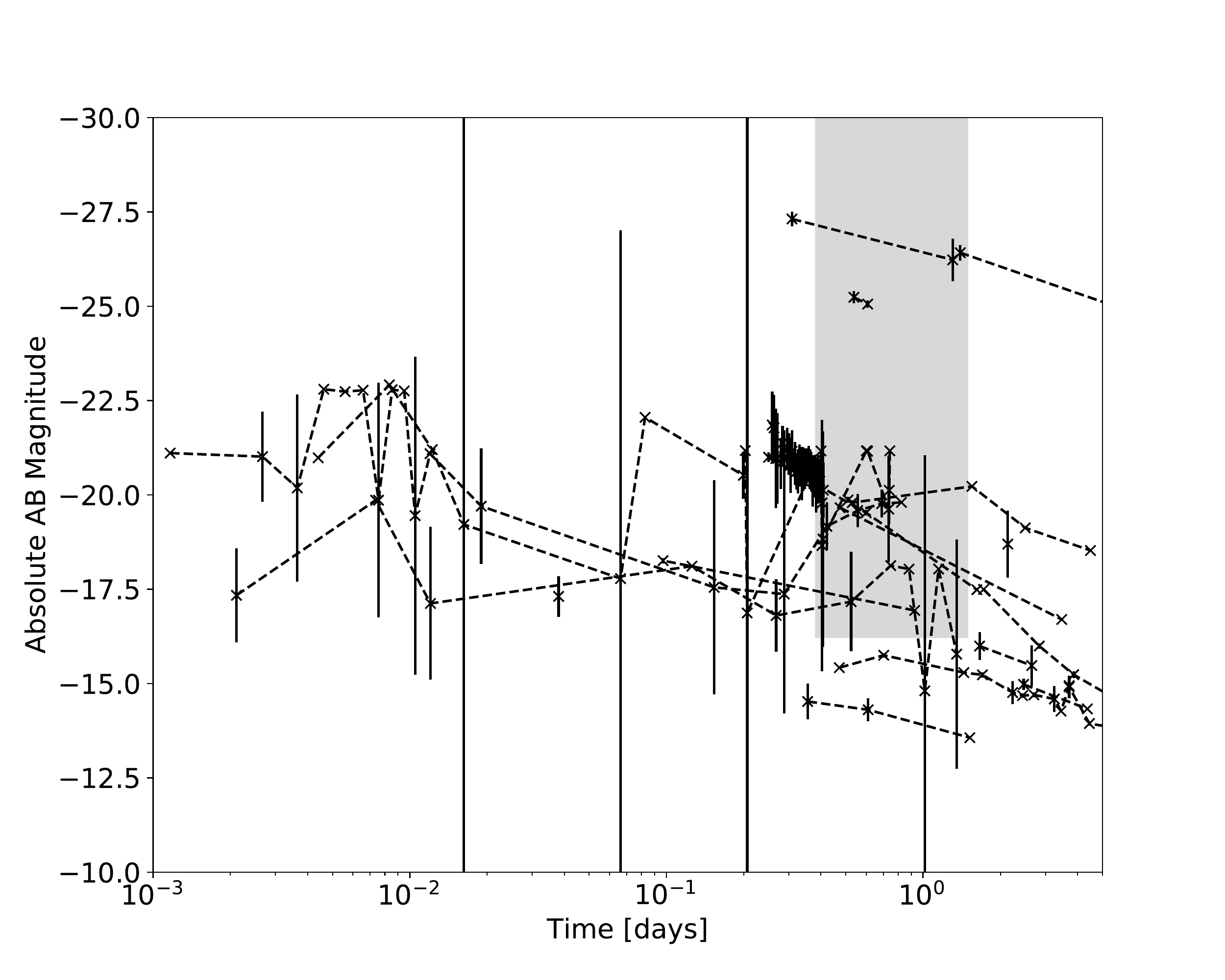}
\end{center}
\caption{The light curves for known SGRBs (in the observer frame) with ZTF limits for comparison. The data are a collection of events with measured redshift from \cite{Fong15}. The ZTF search occupies the grey region in the magnitude versus time after the trigger diagram. On the left, we plot apparent magnitude, and on the right, absolute magnitude, where we have taken a fiducial distance of 200\,Mpc for the SGRB, consistent with the LIGO/Virgo detection horizon \citep{AbEA2018}. This distance was used to transform the ZTF apparent magnitude limits into absolute magnitudes, as well as a revised distance for the SGRBs used in the sample.}

 \label{fig:lightcurves}
\end{figure*}

Models and previous SGRBs detected have characterized the evolution of these kind of transients as rapidly fading sources. Thus our follow-up can discriminate between SGRBs and other transients by taking the difference between the magnitude at the discovery and the magnitude of the follow-up, as well as by examining the lightcurve or the spectrum if available. 
After having observed a significant portion of the probability region and classifying 11 of the 14 remaining transients without finding any potentially viable counterpart for the SGRB, it is possible to compare the search sensitivity, both in terms of depth and timescale, to expected counterpart properties. 
In Figure \ref{fig:lightcurves}, we show where the median limits fall relative to known afterglows.
Over 61.6\% of the localization, we detected no afterglow emission to a depth of $m_\textrm{AB} > 20.3$ in r-band from 0.6 to 2 days after the merger.
This comparison shows that future follow-ups would benefit both from a more rapid response (although in this case, the follow-up was restricted by the day time as opposed to technical issues), and taking longer exposures, resulting in deeper observations (unless the SGRB is in the local universe, as will be the case for the gravitational-wave counterpart searches).
Earlier observations would not only allow for observations of objects when they are brighter, but also allow for easier identification of rapidly fading afterglows, especially when near the sensitivity limit for the telescope.

\section{Conclusions}
\label{sec:conclusions}

Our analysis demonstrates the feasibility of large scale TOO coverage of large sky areas with ZTF. Although we did not find the optical counterpart to the short GRB, the significant coverage of the sky localization region gives confidence in success for future events.
In particular, the lack of unrelated sources, enabled by high cadence pre-burst limits from the regular ZTF survey, makes follow-up of thousands of square degrees possible.
This is a significant improvement upon the intermediate Palomar Transient Factory searches for optical counterparts to long gamma-ray bursts, where for example, a 71 degrees squared search for GRB 130702A was performed in 10 fields, yielding 43 candidates \citep{SiCe2013}.
It also serves as a path-finder for optical follow-up for future gravitational-wave events, which is important after the significant success of GW170817, the first joint GW-EM detection \citep{2017ApJ...848L..21A,ChBe2017,CoBe2017,2017Sci...358.1570D,2017Sci...358.1565E,2017ApJ...848L..25H,2017Sci...358.1579H,KaNa2017,KiFo2017,2017ApJ...848L..20M,2017ApJ...848L..32M,NiBe2017,2017Sci...358.1574S,2017Natur.551...67P,SmCh2017,2017Natur.551...71T,2017PASJ...69..101U}. Some future short GRB detections are likely to yield kilonovae as well \citep{AsCo2018}, which enable joint gravitational-wave and electromagnetic studies \citep{CoDi2018b,CoDi2018}.
Our results show that for kilonovae produced from GW170817, which has a peak absolute magnitude of about $m_\textrm{AB} \sim -16$, would be detectable to about 200\,Mpc, consistent with the LIGO/Virgo detection horizon \citep{AbEA2018}.
While the coverage of 2900 square degrees yielded a coverage of $\approx$\,62\,\% of the probability region, this areal coverage would be more than sufficient to cover greater than 90\% of the probability region for most future gravitational-wave events \citep{AbEA2018}.
We checked that when estimating the distribution of the probability content in sky areas of similar size to this skymap using example gravitational-wave sky maps, on average about 13-15\,\% of the sky area is lost due to chip-gaps.

Going forward, we will be prioritizing minimizing the time between the notice of the short GRB and the imaging by ZTF. 
These observations were scheduled by-hand \citep{CoEA2018c}, and this scheduling has now been fully automated and has been demonstrated on the most recent SGRBs that we have targeted \citep{CoEA2018a,CeEA2018a,CoEA2018b,CoEA2018d}.
Also, there is now a filter for objects without previous detections.
In addition, the scheduling will be done differently, emphasizing images of higher probability fields while also accounting for rising and setting fields and the need to make repeated exposures, potentially in different filters, to maximize the probability of counterpart detection \citep{CoTo2018}.

\acknowledgments

Based on observations obtained with the Samuel Oschin Telescope 48-inch and the 60-inch Telescope at the Palomar Observatory as part of the Zwicky Transient Facility project. Major funding has been provided by the U.S National Science Foundation under Grant No. AST-1440341 and by the ZTF partner institutions: the California Institute of Technology, the Oskar Klein Centre, the Weizmann Institute of Science, the University of Maryland, the University of Washington, Deutsches Elektronen-Synchrotron, the University of Wisconsin-Milwaukee, and the TANGO Program of the University System of Taiwan.
MC is supported by the David and Ellen Lee Postdoctoral Fellowship at the California Institute of Technology.
SG is supported by the National Science Foundation with NSF Award PHY-1607585.

\bibliographystyle{aasjournal}
\bibliography{references}

\begin{sidewaystable}
\centering
\caption{Follow-up for the candidates. $\Delta m / \Delta t$ measures the fade in magnitudes between the first observation with ZTF and the final observation with the follow-up telescopes. For objects with this metric, the lack of rapid measured fading indicates they are unlikely to be associated with the SGRB.}
\label{table:followup}
\begin{tabular}{|l|l|l|l|l|l|l|}
\hline
Candidate &Coordinates & magnitude at discovery& Date of last observation & Data available &Classification&$\Delta m / \Delta t$           \\ \hline
ZTF18aawozzj  & 	12:31:09.02 +57:35:01.8&g=20.2& June 9 &  P200+DBSP Spectrum & SNIa at z=0.095  \\ \hline
ZTF18aawnbgg  &  	10:40:54.05 +23:44:43.3&r=19.88&June 9 &  P200+DBSP Spectrum & SNIa at z=0.13 \\ \hline
ZTF18aawmvbj  &10:12:41.17 +21:24:55.5&r=19.75& June 9 &  P200+DBSP Spectrum & SNIa at z=0.14  \\ \hline
ZTF18aawcwsx  & 	10:40:33.46 +47:02:24.4&r=19.84& June 5 &  P60+SEDM Spectrum & SNIa at z=0.09 \\ \hline
ZTF18aawnbkw  &10:38:47.66 +26:18:51.8&r=19.91& June 12 &KPED r=20.01& & -0.1 mag /20 days \\ \hline
ZTF18aawmqwo  &09:52:06.90 +47:18:34.8&r=19.98& June 21 & KPED r=19.96&& 0.02 mag /20 days\\ \hline
ZTF18aawmkik  & 08:51:11.45 +13:13:16.7& r=19.04&June 12 & KPED r=20.6 && -1.56 mag /20 days\\ \hline
ZTF18aawnmlm  &11:03:11.38 +42:07:29.9&r=20.12& June 19 & KPED r=20.2&&-0.08 mag /18 days\\ \hline
ZTF18aauhzav  & 	10:59:29.32 +44:10:02.7&r=19.97& June 19 &KPED r=21.4 && -0.1 mag /20 days \\ \hline
ZTF18aavrhqs  &11:58:09.57 +63:45:34.6&r=19.99& June 21 & KPED r=21.4&& -1.41 mag /21 days\\ \hline
ZTF18aawmwwk  & 	10:35:26.51 +65:22:34.3&r=19.9& June 21 &KPED r=19.8 & &+0.1 mag /21 days\\ \hline
ZTF18aawwbwm  & 08:16:44.98 +35:34:13.1& r=19.79&Not observable  & &							\\ \hline
ZTF18aawmjru  & 	08:39:11.39 +44:01:53.6&r=18.43& Not observable && & \\ \hline
ZTF18aawmigr  & 	08:48:01.76 +29:13:51.9&r=19.63& Not observable & & \\ \hline

\end{tabular}
\end{sidewaystable}


\end{document}